\documentclass[aps,pra,twocolumn,superscriptaddress]{revtex4-1}

\usepackage{amsmath,amssymb,graphicx}
\usepackage{color,times}
\usepackage{xcolor}
\usepackage{placeins}
\usepackage{siunitx}

\definecolor{darkblue}{rgb}{0, 0, 0.8}
\definecolor{darkred}{rgb}{0.5, 0, 0.0}

\usepackage[colorlinks=true, breaklinks=true, linkcolor=darkblue, citecolor=darkblue, urlcolor=darkblue]{hyperref}
\newcommand{\doilink}[2]{\href{http://dx.doi.org/#1}{#2}}

\long\def\/*#1*/{}

\begin{document}

\title{Enhanced atom-by-atom assembly of arbitrary tweezers arrays}

\author{Kai-Niklas~Schymik, Vincent~Lienhard, Daniel~Barredo, Pascal~Scholl, Hannah~Williams, Antoine~Browaeys and Thierry~Lahaye}
\affiliation{Universit\'e Paris-Saclay, Institut d'Optique Graduate School,\\
CNRS, Laboratoire Charles Fabry, 91127 Palaiseau, France}

\date{\today}

\begin{abstract}
We report on improvements extending the capabilities of the atom-by-atom assembler described in [\doilink{10.1126/science.aah3778}{Barredo \emph{et al.}, Science {\bf 354},  1021 (2016)}] that we use to create fully-loaded target arrays of more than $100$ single atoms in optical tweezers, starting from randomly-loaded, half-filled initial arrays. We describe four variants of the sorting algorithm that (i) allow decrease the number of moves needed for assembly and (ii) enable the assembly of arbitrary, non-regular target arrays. We finally demonstrate experimentally the performance of this enhanced assembler for a variety of target arrays.
\end{abstract}

\maketitle

Over the last few years, arrays of single laser-cooled atoms trapped in optical tweezers have become a prominent platform for quantum science, in particular for quantum simulation~\cite{Browaeys2020}. They allow single-atom imaging and manipulation, fast repetition rates, and high tunability of the geometry of the arrays. When combined with excitation to Rydberg states, these systems naturally implement quantum spin models, with either Ising \cite{Labuhn2016,Bernien2017,Lienhard2018,Kim2018,Keesling2019} or XY \cite{deLeseleuc2019} interactions. They can also be used to realize quantum gates with fidelities approaching those of the best quantum computing platforms \cite{Levine2018,Levine2019,Graham2019,Madjarov2020}.  

A crucial ingredient of the atom array platform is the atom-by-atom assembly of fully-loaded arrays, starting from the partially-loaded arrays (with a typical filling fraction of 50\% to 60\%) obtained when loading optical tweezers with single atoms~\cite{Schlosser2001}. This technique, first demonstrated in \cite{Barredo2016,Endres2016,Kim2016}, can follow different approaches. A fast and effective approach for realizing one-dimensional chains uses an acousto-optic deflector (AOD) driven with multiple radio-frequency tones to generate all the traps~\cite{Endres2016}; after loading, empty traps are then switched off and the remaining ones are brought to their target position, thus achieving  a fully-loaded chain in a single step. However, directly extending this approach to more than one dimension is challenging~\cite{Brown2019}. A different approach consists in using a spatial light modulator (SLM) to generate arbitrary patterns of traps in 1, 2 or 3 dimensions, load them with atoms, and then dynamically  change the SLM pattern to rearrange the atoms in space~\cite{Lee2016_3d}. However SLMs are slow, making the rearrangement time prohibitive, which limits this approach to small atom numbers. Another approach is using a static trap array and combining it with a moving tweezers \cite{Barredo2016,deMello2019}.

Our approach~\cite{Barredo2016} uses an SLM that produces a user-defined fixed pattern of optical tweezers which includes the  final (target) array, combined with a moving tweezers. This extra microtrap, controlled by a 2D-AOD, is used to move the atoms one by one to reach a fully-loaded target array. The heuristic `shortest-moves-first' algorithm used in \cite{Barredo2016} to find the set of needed moves is versatile, as any target array included in an initial regular array can be assembled. It works well up to a few tens of atoms, but it has some limitations. Firstly, the algorithm was written for regular arrays, such as square and triangular lattices. On completely arbitrary arrays, lattice edges along which atoms can be moved are not naturally given, and using straight paths between source and target traps would lead to unwanted losses, as another target trap already containing an atom may be in the way. Another limitation is that the number of moves needed for ordering is not optimal, and minimizing this number becomes more crucial when the number $N$ of assembled atoms increases beyond a few tens. 

Here, we describe four improved algorithms that can be used without any change in the hardware; the choice of the most efficient approach depends on the characteristics of the target array. We first recall in Sec.~\ref{sec:pb} the problem we need to solve, and review our previous approach and its shortcomings (Sec.~\ref{sec:old}). We then discuss in Sec.~\ref{sec:compress} a compression algorithm which is well adapted for compact arrays (here, by compact, we mean that no trap other than target ones lie within the target array). The number of moves is then at most $N$, which significantly reduces the assembly time. We show in \ref{sec:lsap} that a similar scaling can be obtained for all arrays (compact or not) by using algorithms based on a Linear Sum Assignment Problem (LSAP) solver. In Sec.~\ref{sec:arb}, we extend these algorithms to the case of fully arbitrary two-dimensional patterns (i.e. that are not embedded in a regular Bravais lattice). Finally, in Sec.~\ref{sec:manip}, we experimentally implement  these approaches in a variety of arrays. 

\section{Statement of the problem}
\label{sec:pb}

Our goal is to obtain a fully-loaded array of $N$ traps, whose positions are given by the user (this defines  the \emph{target} array; denoted by green circles in this paper). To do so, we start from a larger array, with $\sim 2N$ traps, containing the target array and extra, \emph{reservoir} traps (these will be denoted by red circles). The entire array is loaded in a stochastic way with a $\sim 50\%$ filling fraction at each realization of the experiment. Therefore we have, with high probability, at least $N$ atoms in the full array. Using a moving optical tweezers, we then transport the atoms one by one, from an initial trap to one of the target traps, until the target array is fully filled. 

To maximize the success probability of the assembly process, we need to minimize the total assembly time. A first reason for that arises from the vacuum-limited lifetime of a trapped atom, which, in our experiments, is $\tau_{\rm{vac}}$ \SI{\sim 20}{\second}. This means that for an array with $N$ atoms, the lifetime of the configuration is $\tau_{\rm{vac}}/N$. It is thus important, when $N$ increases, to minimize the total assembly time to reduce atom losses during rearrangement. As atoms are moved between traps at a constant velocity (typically \SI{\sim 100}{\nano\meter\per\micro\second}, meaning we need \SI{\sim 50}{\micro\second} to move over a typical nearest-neighbor distance of \SI{5}{\micro\meter}), and as it requires a comparatively longer time (\SI{600}{\micro\second}) to capture or release an atom~\cite{Barredo2016}, minimizing the arrangement time mainly amounts to minimizing the number of moves and, but to a lesser extent, the total travel distance. A second reason for minimizing the number of moves is that each transfer from a source trap to a target trap has a finite success probability $p$ (typically around $p\sim0.98-0.99$ in our experiments), partly due to the already mentioned vacuum-limited losses, but also due to, e.g. inaccuracy in the positioning of the moving tweezers, or residual heating. Beyond the number of moves and the total travel distance, the time it takes for the algorithm to compute the moves at each repetition of the experiment contributes to the total assembly time.

In \cite{Barredo2016}, we distinguished two types of moves for reordering. The first approach (that we called type-1) corresponds to the situation where the atom can be moved in between adjacent rows of traps. Then, as on average  $N/2$ atoms are out of place initially, the mean number of needed moves is $N_{\rm mv}=N/2$, and we have to solve a linear sum assignment problem \cite{IntroAlgo}. Using the  Hungarian algorithm (as in \cite{Lee2016_hung}) then minimizes the assembly time. However, type-1 moves require a large distance (at least \SI{5}{\micro\meter}) between any two traps, to avoid atom loss due to disturbances of the trap potential. In practice, many experimental reasons (the finite field of view of the lenses used to focus the tweezers, the need to have large interaction strength between Rydberg atoms, and to have uniform Rydberg excitation lasers over the array) call for having smaller distances in the arrays. Furthermore, as we shall see in Sec.~\ref{sec:arb}, type-1 moves are not well suited for the assembly of truly arbitrary geometries. For these reasons, we here focus on solving our problem using just type-2 moves, where an atom can only be moved along a straight path between adjacent traps. 

In the case of type-2 moves, assigning any source trap to any target trap is not possible, since other atoms might be in the way and need to be moved first. Finding the optimal set of moves is nontrivial since it requires finding a collision-free assignment with a well-defined ordering of the moves. In computer science, this problem is known as the ``pebble motion on a graph'' (in a variant with unlabeled pebbles), and is intractable for large $N$~\cite{Calinescu2006}, even more so in practice as we need to solve it in a time short compared to the configuration lifetime. Therefore, we opt for heuristic algorithms, provided they give a solution not too far from the optimum, and run in a few tens of ms at most for up to a few hundreds atoms. In the next section, we will see that the algorithm used in \cite{Barredo2016} actually meets these criteria only when the target array is not too compact, and when $N$ is not too large. 

\section{Our previous assembler: principle of operation and limitations}
\label{sec:old}

\begin{figure}[t]
\centering
\includegraphics[width=85mm]{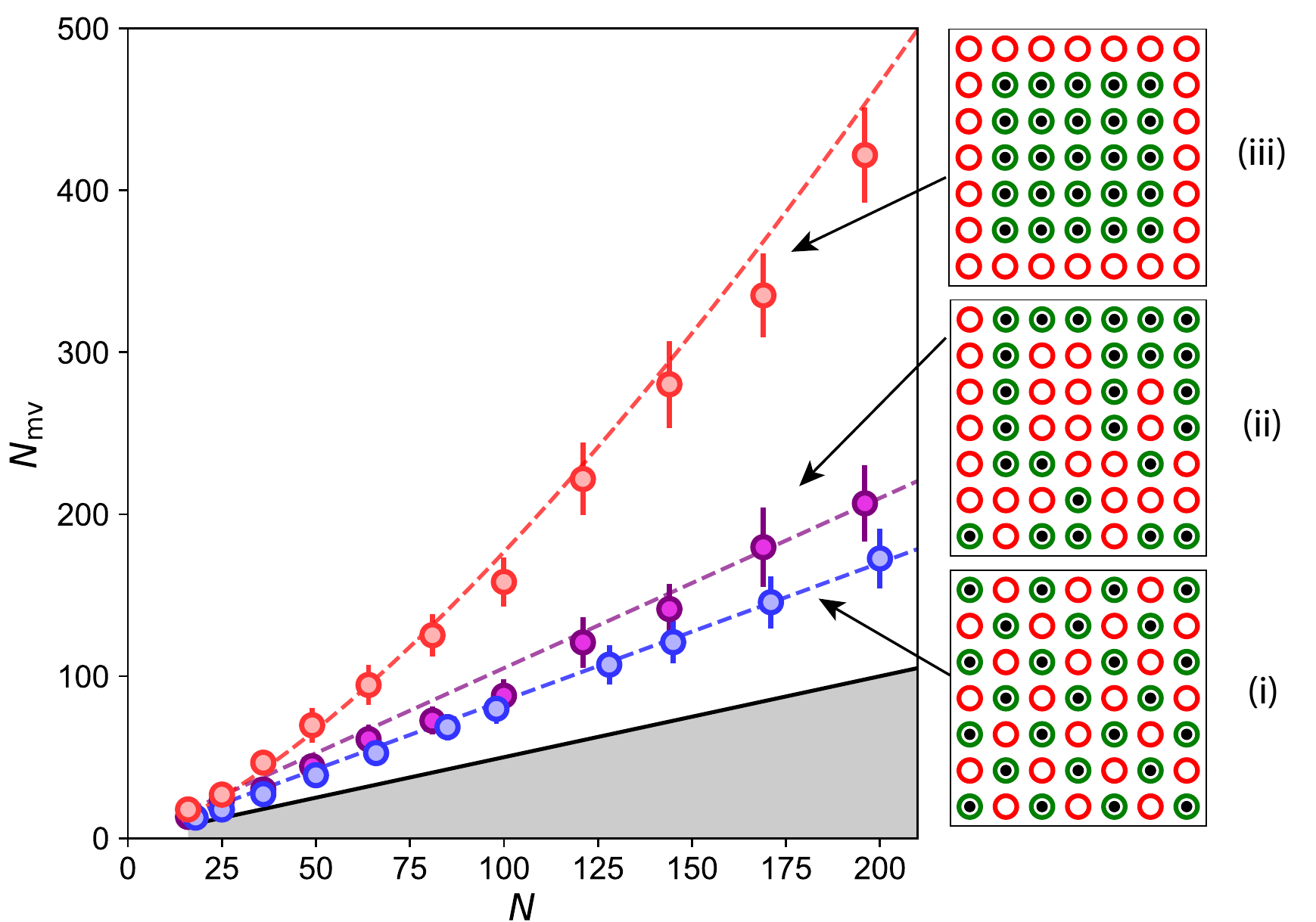}
\caption{\textbf{Scaling of the number of moves for different geometries, with the ``shortest-moves-first'' approach.} The plot shows the number $N_{\rm mv}$ of moves (averaged over  1000 realizations of the random loading; the error bars indicate the standard deviation of the distribution of $N_{\rm mv}$) as a function of the size $N$ of the target array.  For staggered configurations (blue), where a target trap and a reservoir trap alternate, the overhead as compared to the lower bound $N/2$ (indicated by the solid black line above the gray-shaded area) is small. For a random subset of target traps in a square array (purple), the number of post-processing moves due to obstacles is already bigger, but the scaling is still linear with $N$. A drastic change appears in the case of compact geometries (red), where the target array is surrounded by reservoir atoms. Here the number of moves does not increase linearly with $N$, but rather as $N^{1.4}$ (dashed line) and many post-process moves are needed. This means that the current algorithm is unsuited for large compact geometries.}
\label{fig:CurrentAlgo_DifferentGeos}
\end{figure}

\begin{figure*}[t]
\centering
\includegraphics[width=0.9\textwidth]{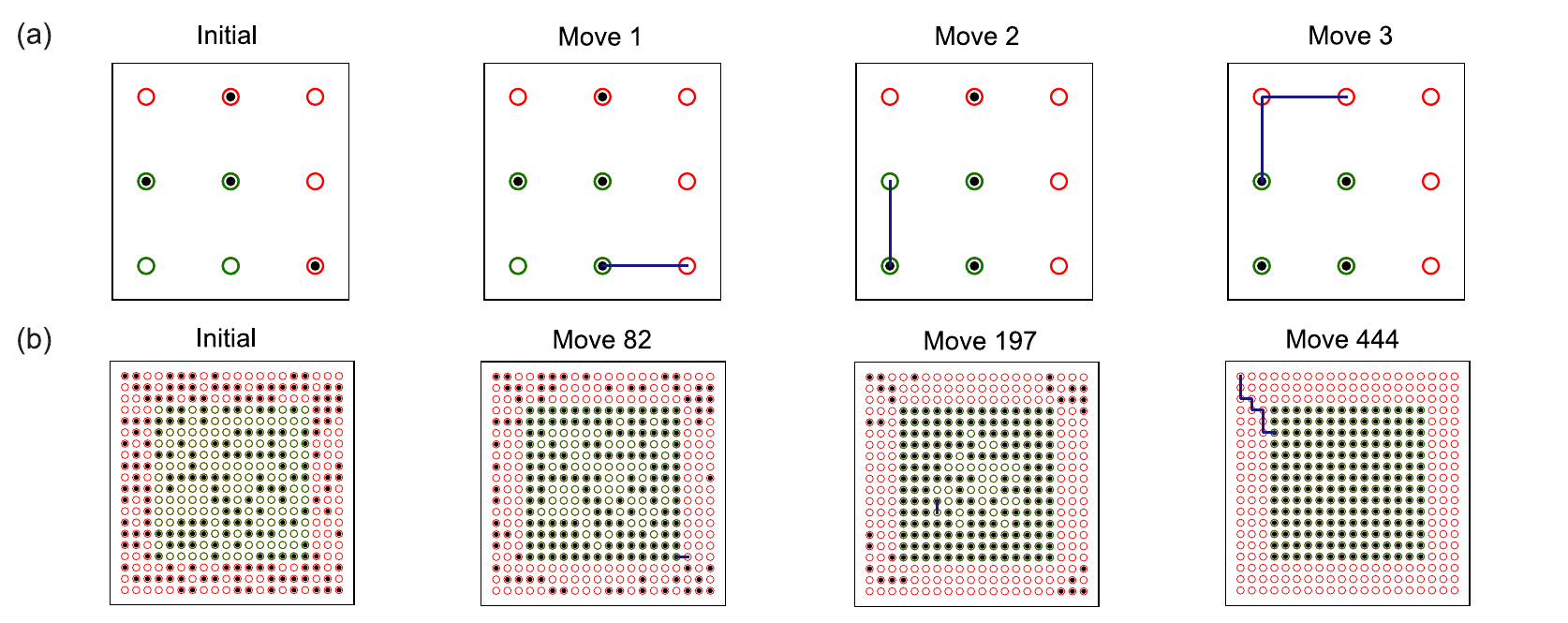}
\caption{\textbf{Assembling of a compact array using the ``short-moves first'' algorithm.} (a) Microscopic view. The first set of moves (blue lines) connects out-of-place atoms with target traps on the outer shell of the structure (e.g. move 1). Once the border is populated, it is no longer possible to fill the inner traps without performing extra moves (move 2). (b) The macroscopic behavior on a $14\times 14$ array reveals that the algorithm starts by filling the border of the target array (green circles) with atoms from reservoir traps (red circles), while inner traps are still empty (e.g. move 82), leading to a large overhead in the number of moves for successful assembling. }
\label{fig:fig2}
\end{figure*}

The atom-by-atom assembler described in \cite{Barredo2016, Barredo2018} allowed us to create user-defined arrays in one, two and three dimensions at unit filling. Non-periodic structures, or complex lattices such as ladders, honeycomb, kagome, or pyrochlore geometries could also be obtained by selecting a subset of target traps on an underlying Bravais lattice. 

We chose a heuristic approach to the problem that had the advantage of requiring a short computation time, scaling as $O(N^2)$, albeit at the expense of not guaranteeing the optimal assignment. This greedy algorithm, that we will call ``shortest-moves-first'', works as follows. We first compute a matrix of distances $D = d_{i j}$ between each out-of-place atom $s_i$ and each (empty) target $t_j$ trap. Then, we order the entries of this matrix by increasing length and select the first $N/2$ elements with the condition that only one element per row or column is chosen (i.e., that each atom or target trap is only assigned once).

This initial matching is not collision-free, as already filled traps may lie in between a matched reservoir atom and an empty target trap. Therefore, in a second step, we post-process this assignment  by applying a rule that splits each move [$S \rightarrow T$] from a source atom $S$ to a target trap $T$ in two moves [$O \rightarrow T$] and [$S \rightarrow O$] for each obstacle atom $O$ that is found in the path. Note that this process leaves the travel distance unchanged, but increases the number of moves, therefore increasing the total assembly time.

Figure~\ref{fig:CurrentAlgo_DifferentGeos} shows the number of moves $N_{\rm mv}$ returned by the above algorithm to assemble a target array of $N$ atoms embedded in a square array, for three different geometries: (i) a checkerboard or staggered pattern, (ii) a random pattern, and (iii) a compact square in the center. The number of moves is averaged over 1000 realizations of the initial random loading. We observe that $N_{\rm mv}$ is only slightly above $N/2$ for the cases (i) and (ii) where reservoir and target traps are strongly mixed. However, in the case (iii) of compact geometries, where all the reservoir atoms lie outside the target array, this procedure scales as $N_{\rm mv}\propto N^\alpha$ with $\alpha\simeq 1.4$ (red dashed line), making it unsuitable for large arrays.

The reason for this is illustrated in Fig.~\ref{fig:fig2}, which shows a few snapshots of the reordering process. The shortest moves are those connecting out-of-place atoms with target traps on the border of the array, therefore the algorithm starts by filling the outermost shell. Once this is done, it is no longer possible to fill the (empty) inner traps without performing extra operations to displace the atoms in the way, giving rise to many extra moves to fill the inner part of the target array. For the initial configuration in Fig.~\ref{fig:fig2}(b), the $14\times14$ target array is assembled in 444 moves. As picking-up and releasing an atom takes extra time, this behavior leads to prohibitive rearrangement times, even if the distance traveled was close to optimal. 

This behavior is problematic, as many arrays of interest for quantum simulation are compact. Therefore, it is crucial to find an assignment between reservoir and target traps which really minimizes the number of moves. For assembling compact arrays, a much better approach, where the maximum number of moves is at most $N$, is the compression algorithm that we now describe.

\section{Improved assembly of compact arrays by the ``compression'' algorithm}
\label{sec:compress}

\begin{figure*}[t]
\centering
\includegraphics[width = 0.9\textwidth]{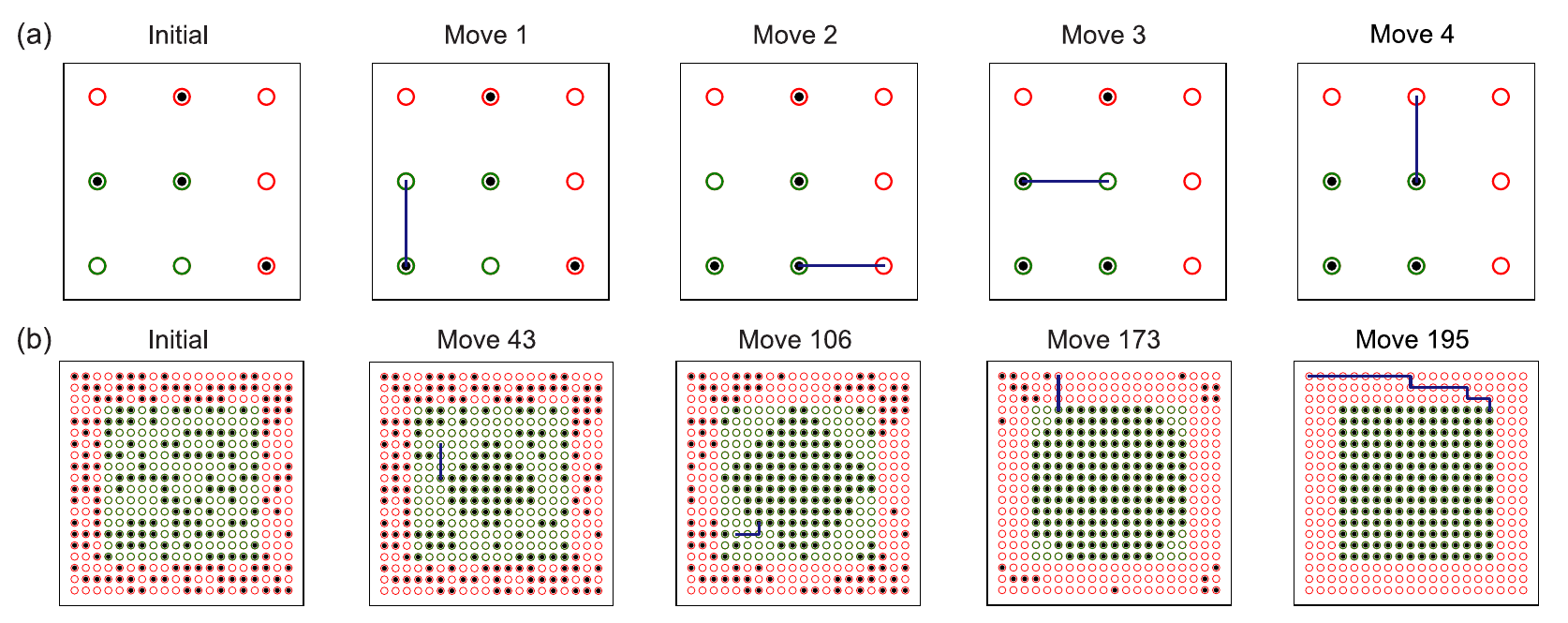}
\caption{\textbf{The compression algorithm.} (a) Illustration of the compression procedure for a $2\times 2$ target sub-array, requiring four moves. (b) A few assembling steps using the compression algorithm to assemble a $14\times 14$ target sub-array in only 195 moves, to be contrasted with the 444 moves needed previously. }
\label{fig:compress_algo}
\end{figure*}

From the above considerations, it is clear that we need to prevent the formation of the outer shell during the assembling process. A simple way to do this and have a collision-free assignment without any post-processing is to  fill the target traps in a determined order. We first fill the central traps, and progressively, one layer after the other, we fill the compact structure until we reach its border. To fill the traps, we choose the closest atoms lying outside the already assembled bulk. An asset of this compression approach is that we can calculate once, independently of the initial loading, a look-up table. The table stores which source traps can be used to fill a given target trap. In combination with the predetermined order in which the target traps are filled, the look-up table reduces the calculation time on a particular instance. It scales roughly as $N^{1.2}$ with the number of target traps and amounts, in our implementation, to about \SI{7}{\milli\second} for $N=100$ on a regular desktop computer with 16~GB of RAM. 

Figure~\ref{fig:compress_algo}(a) illustrates how the algorithm works on a small square array. The target array is first assembled from the bottom left corner, then the diagonal, and finally the top right corner. Using this algorithm, atoms which initially occupy target traps can be displaced, which  means additional moves with respect to an optimal solution. But, as we always use the closest atoms to the border of the compact structure to assemble it, the path is always obstacle-free and therefore we do not need any post-processing. Consequently, each atom is moved \emph{at most once} during the assembling process, which sets the  upper bound $N_{\rm mv}\leq N$ and ensures on average a smaller number of moves using the compression algorithm with respect to the shortest-moves-first algorithm of the previous section. As $N_{\rm mv}$ can not be lower than $N/2$ on average, our solution, while not optimal for many initial loading instances, is generally close-to-optimal. Figure~\ref{fig:compress_algo}(b) shows how this compression algorithm outperforms the shortest-moves-first one. The 196 target atoms are assembled in 195 moves, whereas the same initial configuration required 444 moves to be sorted with our previous approach.

\begin{figure}[b]
\centering
\includegraphics[width=85mm]{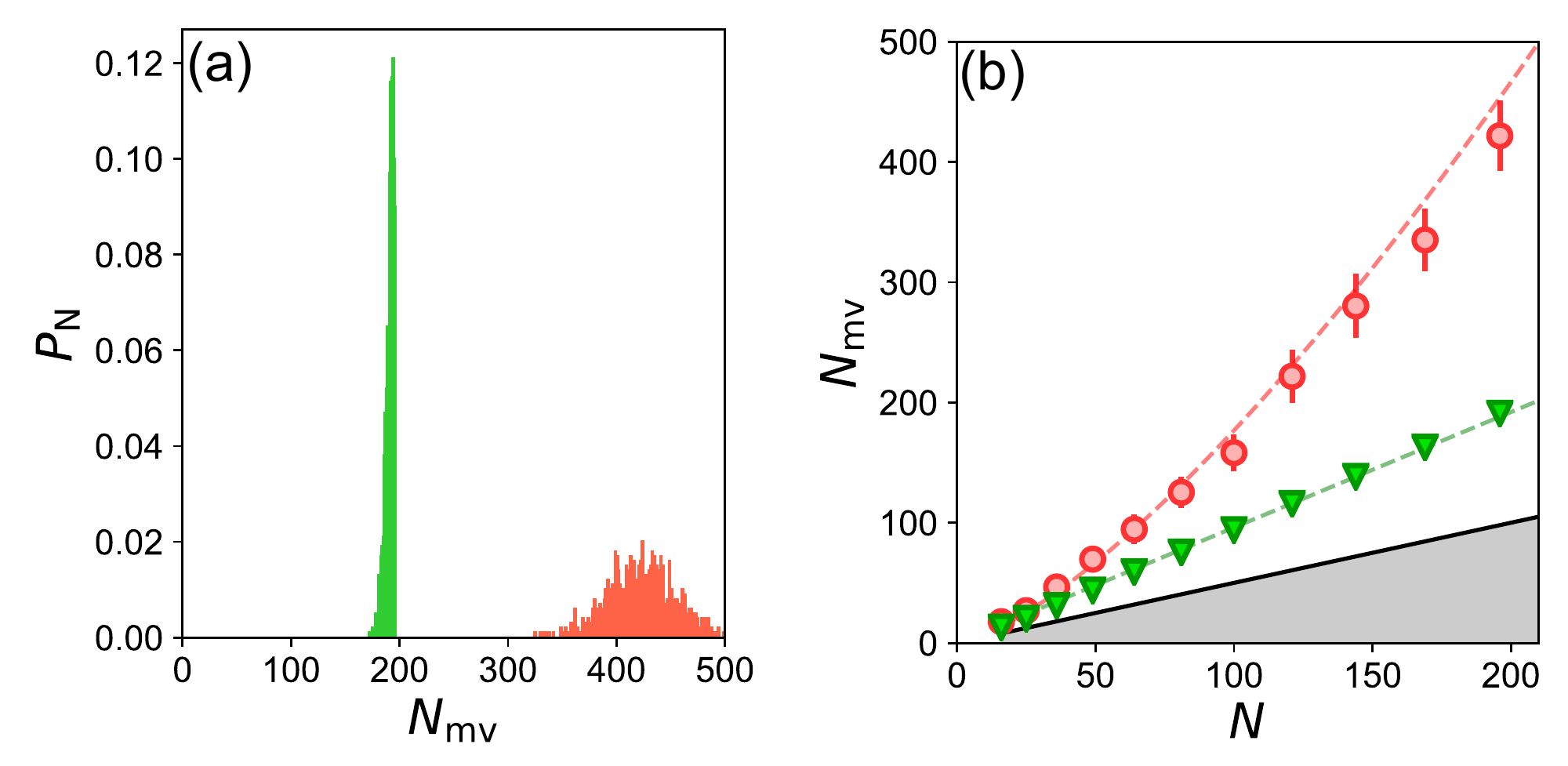}
\caption{\textbf{``Compression'' vs. ``Shortest-moves-first'' algorithms}. (a) Histogram of the number of moves needed to fill a $14\times 14$ square array for 1000 initial random loading instances. The compression algorithm (green) has a narrow distribution which is bounded by $N$. The shortest-moves-first algorithm (red) has a broad distribution and requires on average many more moves since the initial assignment is not collision-free. (b) Comparison of the scaling of $N_{\rm mv}$ as a function of $N$ between the two algorithms. The compression algorithm gives a number of moves linear in $N$. Error bars are the standard deviation of the distribution. }
\label{fig:hist_compress_algo}
\end{figure}

As can be seen in Fig.~\ref{fig:hist_compress_algo}(a), not only is the average number of moves smaller than before, but the distribution of $N_{\rm mv}$, calculated for 1000 random initial loading instances of the array, is also strongly sub-Poissonian, and asymmetric, with a sharp cutoff at $N$. This is an appealing feature, as it indicates that the success probability of the assembly process should be more consistent from one shot to the other, as compared to the previous approach. Figure~\ref{fig:hist_compress_algo}(b) shows the linear scaling of $N_{\rm mv}$ with $N$.   

This technique can be naturally extended to the case of compact structures in other lattices (e.g. triangle) and also to arbitrary geometries, as we shall see in section~\ref{sec:arb}.

\section{Using a Linear Sum Assignment Problem solver}
\label{sec:lsap}

In view of minimizing the number of moves, it is interesting to revisit the approach of the problem as a Linear Sum Assignment Problem (LSAP), that was mentioned above for the case of type-1 moves. However, for the type-2 moves we are interested in here, a direct application of the LSAP matching with the  travel distance $\ell$ as a cost function does not yield a collision-free assignment and requires post-processing, which in general increases the number of moves. We describe in this section two different algorithms, that first use a LSAP solver and then reprocess the moves, which lead to a low number of moves. The LSAP solver we use in practice is a modified Joncker-Volgenant algorithm with no initialization \cite{LSAP_Scipy} which is implemented in the scipy.optimize Python package \cite{2020SciPy-NMeth}. 

\begin{figure*}[t]
\centering
\includegraphics[width=16cm]{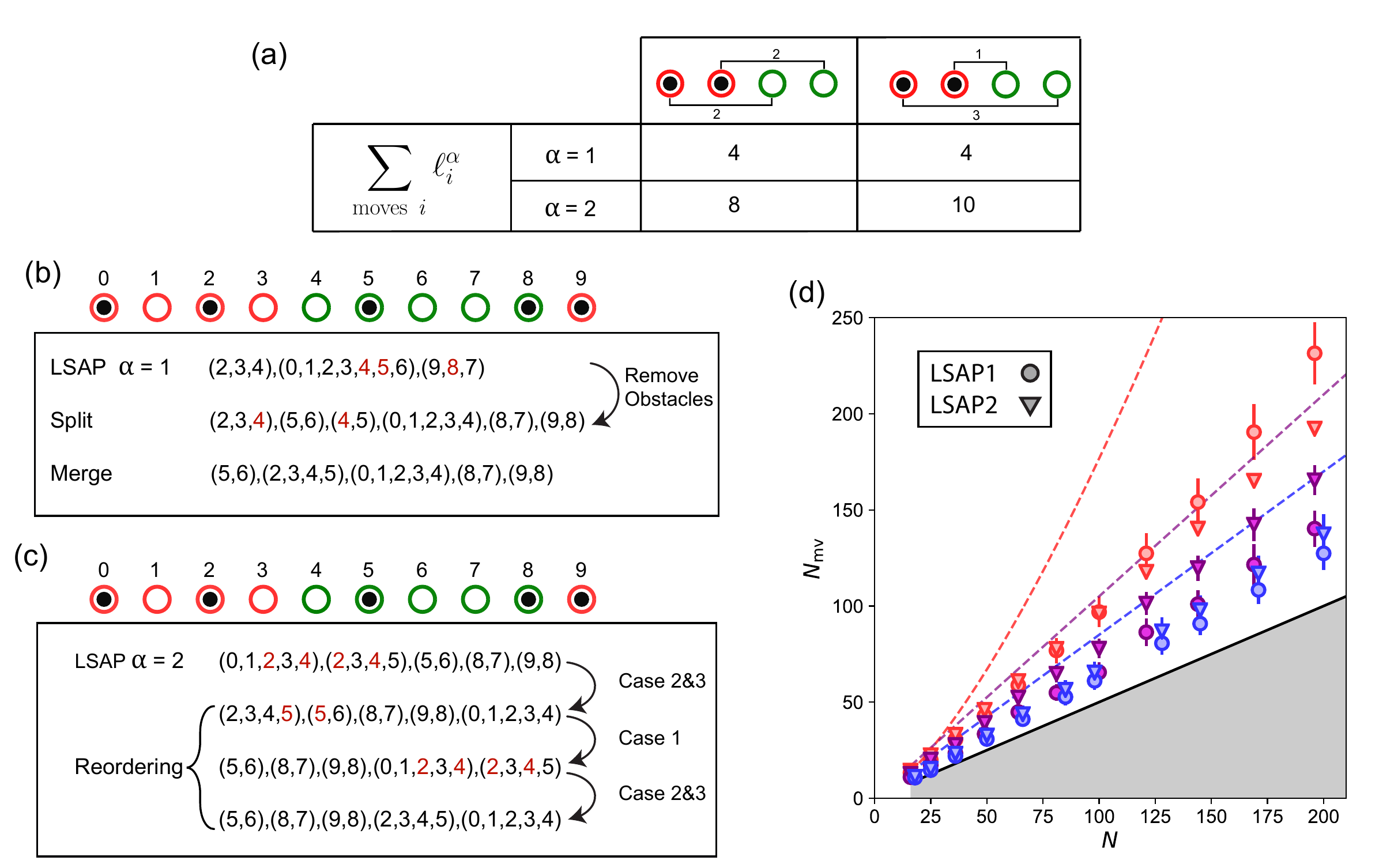}
\caption{{\bf Modified LSAP algorithms.} (a) Using a cost function with $\alpha=2$ (see text) in a LSAP solver favors short moves. (b) The algorithm LSAP1 first uses a LSAP solver with $\alpha=1$, which returns a list of moves (here, $(2,3,4)$ means that the atom initially in trap 2 is moved, via trap 3, to trap 4). Some moves lead to collisions (denoted in red) and thus the set of moves is post-processed as in the ``shortest-moves-first'' algorithms, by splitting the problematic moves into two or more stages. However, in a second step, two moves that share the same trap as final and initial positions (denoted in red) can be merged together, reducing the total number of moves. (c) The algorithm LSAP2 uses a modified cost function with $\alpha=2$, which returns a set of short moves; to avoid collisions, the moves are then reordered by applying successively three rules (see text) until the rearrangement can be performed without collisions. Numbers in red highlight the breaking of a rule. (d) Number $N_{\rm mv}$ of needed moves as a function of $N$ to assemble a checkerboard subarray (blue), a random subarray (purple) or a compact subarray (red), for the LSAP1 and LSAP2 algorithms. The dashed lines reproduce the fits of Fig.~\ref{fig:CurrentAlgo_DifferentGeos} for comparison. }
\label{fig:LSAPs}
\end{figure*}

The first algorithm (LSAP1) uses the total travel distance $\sum_{\rm moves\;\textit{i}}\ell_i$ as the cost function, while the second one (LSAP2) uses a modified metric $\sum_{\rm moves\;\textit{i}}\ell_i^2$, which favors shorter paths (Fig.~\ref{fig:LSAPs}(a)). In both cases, the set of returned moves is post-processed to eliminate collisions and reduce the number of moves. 

\subsection{LSAP1: standard metric and merging}

Our first approach, described on a simple example in Fig.~\ref{fig:LSAPs}(b), starts with the LSAP algorithm using the travel distance between source and target traps as a cost function. We first sort the returned moves from  shortest to longest. Since the found assignment leads to collisions, we then post-process the set of moves by splitting the paths with obstacles into two or more moves, just as in the shortest-moves-first approach. However, in a second iteration, we merge again some moves in which an atom is picked-up twice, thereby reducing the number of moves considerably, checking at each step that we do not reintroduce any collision in doing so. Note that this second merging iteration can in principle be applied to any algorithm, but yields the smallest number of moves when starting from the LSAP matching. The computation time for this approach is on average 4 ms for 100 target traps in a staggered geometry, and roughly scales as $N^2$~\cite{note-LSAP-Scaling, note-Lookup-table}.

Figure~\ref{fig:LSAPs}(d) shows the number of moves $N_{\rm mv}$ as a function of $N$ for LSAP 1 (disks). The performance is very satisfactory for staggered or random target arrays, as the number of moves is only 20 to 30~\% higher than the absolute lower bound $N/2$. For compact arrays, the number of needed moves is slightly larger than $N$, making this approach less efficient than the compression algorithm described in \ref{sec:compress}.

\subsection{LSAP2: modified metric and reordering}

Long moves lead to many collisions, therefore it is beneficial to avoid them. In our second approach we achieve this by using a modified cost function $\sum_{\rm paths\;i}\ell_i^2$. A similar idea was introduced in~\cite{Lee2016_hung}, but here the moves are sequential, and we thus need to find the right ordering in which the moves have to be performed to avoid collisions.

To do so, we apply the following rules. We examine each move in the list, and, if the target trap of the move is occupied (case 1), or if another trap along the path of the move is filled (case 2), or if the target trap is in the path of another move following in the list (case 3), we postpone this move and put it at the end of the list of moves. We find empirically that this procedure always produces a collision-free set of moves. This approach is illustrated in Fig.~\ref{fig:LSAPs}(c). The whole algorithm (LSAP and reordering) has an average computation time of 4~ms for $N=100$ target traps in a compact geometry, and scales roughly  as $N^2$.

Whatever the target array, the maximum number of moves is bounded by $N$, the size of the cost matrix. As can be seen in  Fig.~\ref{fig:LSAPs}(d) (triangles), the number of moves returned by LSAP2 is slightly larger than LSAP1 for sparse arrays, but is smaller for compact arrays, where it gives essentially the same performance as the compression algorithm. The latter, however, has the advantage of a shorter calculation time for $N>N_{\rm c}$, with, in our current implementation, $N_{\rm c}\sim 300$.

\section{Arrays with completely arbitrary geometry}
\label{sec:arb}

\begin{figure}[t]
\centering
\includegraphics[width=85mm]{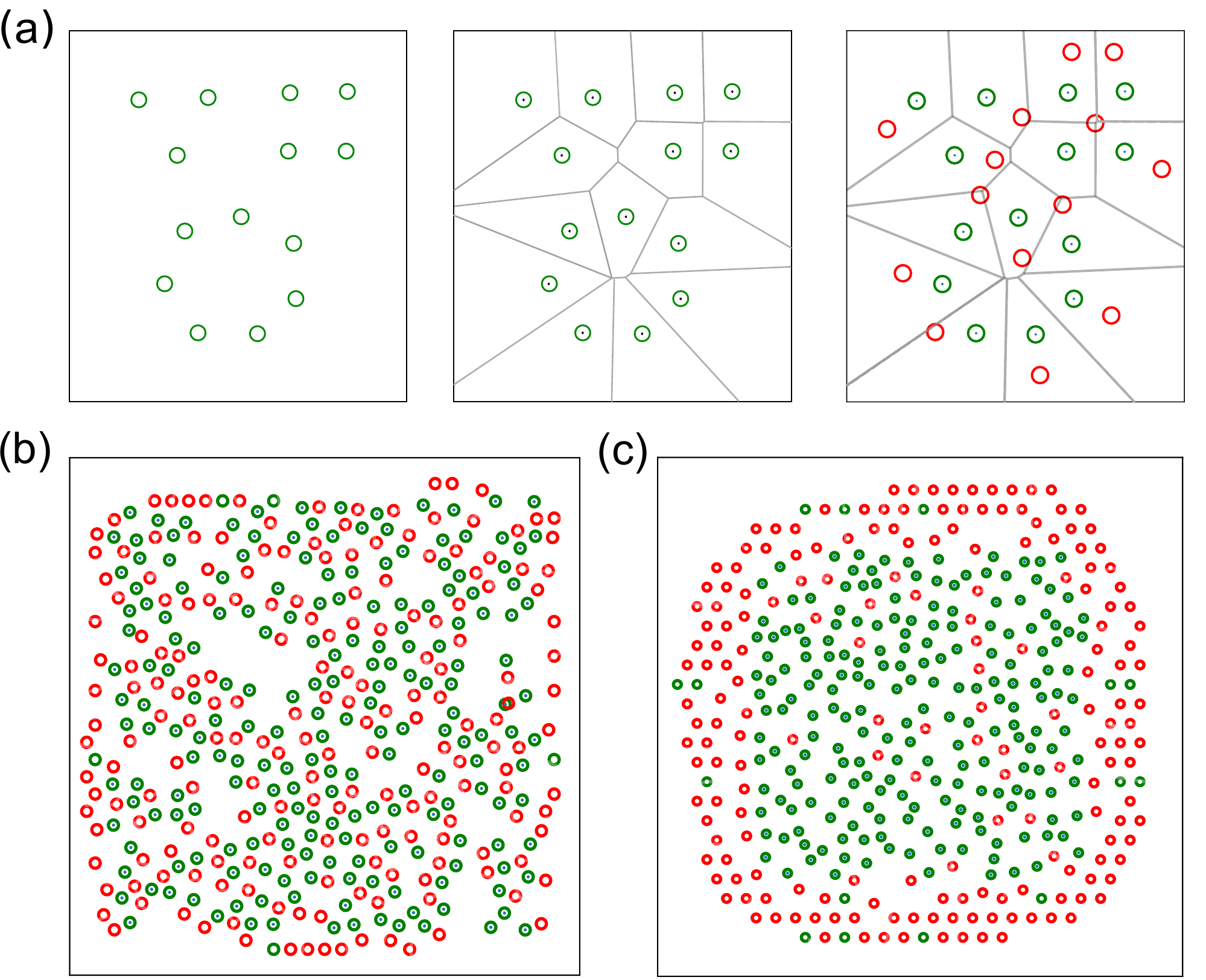}
\caption{\textbf{Generating the reservoir arrays for arbitrary target arrays.} (a) Starting from the user-defined target array (left), we compute its Voronoi diagram (middle) and, in each cell we add a reservoir trap, shown in red, if there is enough room (right), otherwise we add it at the periphery (see text for details). (b,c) Examples of generated reservoirs for a $N=200$ target array, without (b) and with (c) the need to add reservoirs at the periphery.}
\label{fig:OptimalReservoir}
\end{figure}

Condensed-matter models are often studied on specific crystalline arrangements which are described by a Bravais lattice, e.g. a square or a triangular lattice. Our previous assembler was therefore based on such an underlying lattice, which simplifies the problem in two ways. Firstly, this naturally defines the paths along which the moving tweezers can travel, and, because these lattice edges are separated by a constant spacing, it automatically ensures that a minimal distance between atoms in traps and the moving tweezers is always kept during the rearrangement. Secondly, it simplifies the distance calculation between two traps by defining the metric in terms of lattice coordinates (Manhattan distance).

Not all physical structures of interest for quantum simulation, however, can be described by a Bravais lattice. Examples of such non-periodic features include crystals with defects (interstitial defects, vacancies, dislocations, grain boundaries), quasi-crystals, disordered arrays for Anderson or many-body localization studies, and even totally arbitrary structures in the context of combinatorial optimization problems such as finding the maximum independent set of a graph \cite{Pichler2018,Henriet2020}. To examine such systems, we developed a variant of our algorithms, which is not based on an underlying lattice, and therefore allows us to assemble truly arbitrary structures.

\begin{figure}[t]
\centering
\includegraphics[width=40mm]{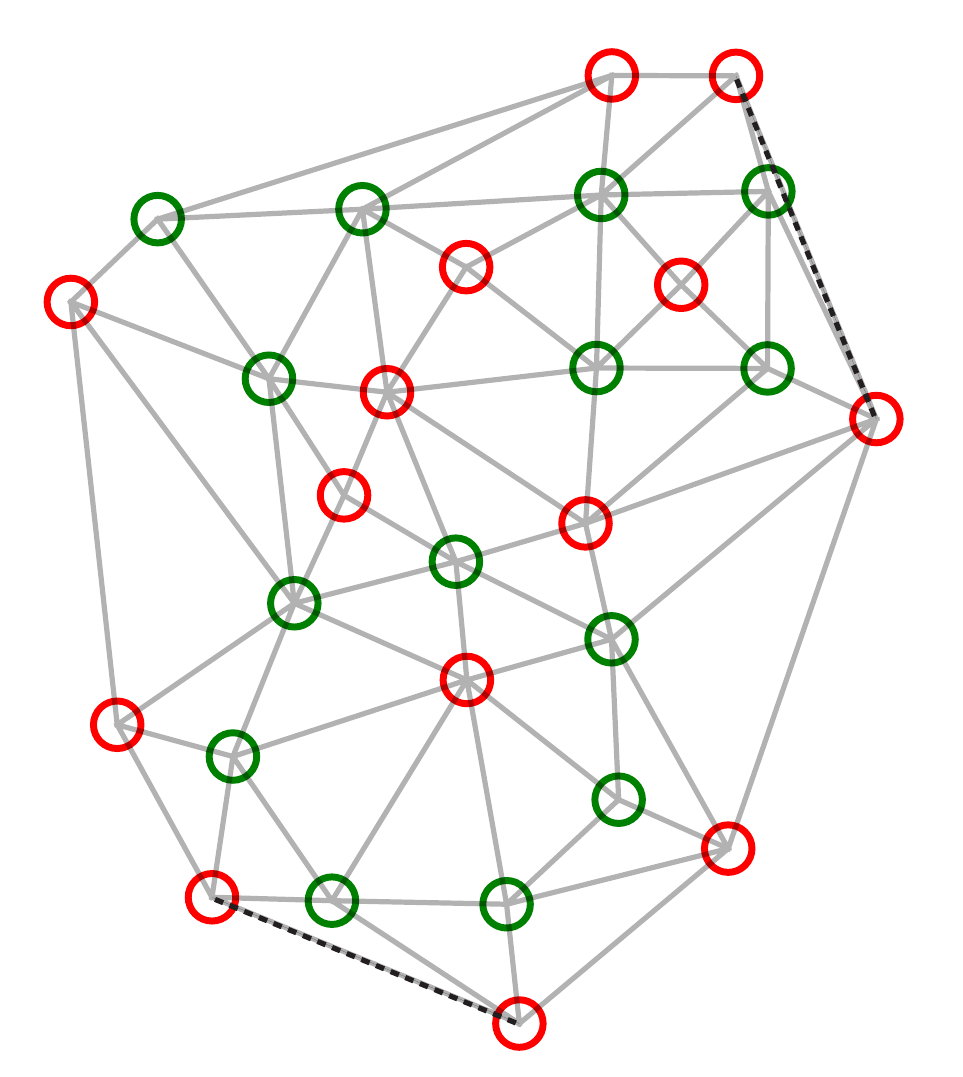}
\caption{\textbf{Generating the allowed paths between traps.} We first perform the Delaunay triangulating the atom array. In a second step, we remove edges which do not fulfill a minimal passing-distance requirements (dotted line).}
\label{fig:FindEdges}
\end{figure}

The starting point for our algorithm is the set of $N$ target traps, whose positions are provided by the user. Because of the stochastic loading, we have to place $N$ additional reservoir traps close to the arbitrary $N$-atom target configuration. This reservoir generation works as follows (Fig. \ref{fig:OptimalReservoir}a). Whenever possible, to reduce the number of moves, a reservoir trap should be placed in immediate proximity of each target trap. To do so, we compute the Voronoi diagram \cite{comp_geom} of the set of target traps (i.e. divide the plane in $N$ regions, one around each target trap $T$, such that all points of this region is closer to $T$ than to any other trap). We then add in each Voronoi cell a single reservoir trap, provided it can be placed at a distance larger than a ``safety'' distance $d_{\rm m}$ (typically \SI{\sim 4}{\micro\meter}) from all other traps. If successful, this procedure ensures that for each target trap there is a single reservoir close to it (Fig. \ref{fig:OptimalReservoir}b). If, however, the density of the target traps is already comparable to $1/d_{m}^2$, then we cannot add enough reservoir traps in this way, and so we place extra traps at the periphery of the pattern in a compact triangular array (Fig. \ref{fig:OptimalReservoir}c). 

The next step is to find paths along which an atom can travel to an empty target trap. Contrary to the case of Bravais lattices, no obvious edges are \emph{a priori} connecting the traps along which the moves can be performed. Direct, straight-line paths from reservoir to target trap are also not possible, since there can be other traps in the way, leading to collisions and atom losses. We thus define the set of allowed paths by using a Delaunay triangulation \cite{comp_geom} of the full set of traps (target and reservoir) as shown in Fig.~\ref{fig:FindEdges}. In practice, we implemented the triangulation in Python 3.0 with the Scipy library \cite{2020SciPy-NMeth}. To enforce the above-mentioned constraint of a minimal passing distance, we post-remove edges that do not meet this requirement (see dashed lines in Fig. \ref{fig:FindEdges}). We emphasize that the generation of the reservoir traps and of the allowed edges are done just once for any given target array, and not at each repetition of the experiment, which considerably relaxes the constraints on the speed of this algorithm. In practice, arrays with hundreds of target traps can be processed in a few seconds.

This triangulation then allows us to naturally describe the whole structure in terms of graph language, connecting the nodes (trap positions) by edges along which the atoms are allowed to move. In this way, we eliminate the necessity to describe the problem with an underlying Bravais lattice. Furthermore, it allows the implementation of efficient shortest-path graph algorithms (e.g. the Dijkstra algorithm \cite{IntroAlgo}) to find the shortest path between a matched initial and target trap, following the allowed edges of the graph. For the generation of the graphs and graph-algorithms the Networkx library \cite{Networkx} is used. With these modifications, it is now possible to extend the algorithms discussed above to arbitrary patterns. The scaling and performance of the algorithms (in terms of computation time and of number of moves) are essentially unchanged as compared to the case of regular lattices.

\section{Experimental demonstration}
\label{sec:manip}

\begin{figure}[t]
\centering
\includegraphics[width=80mm]{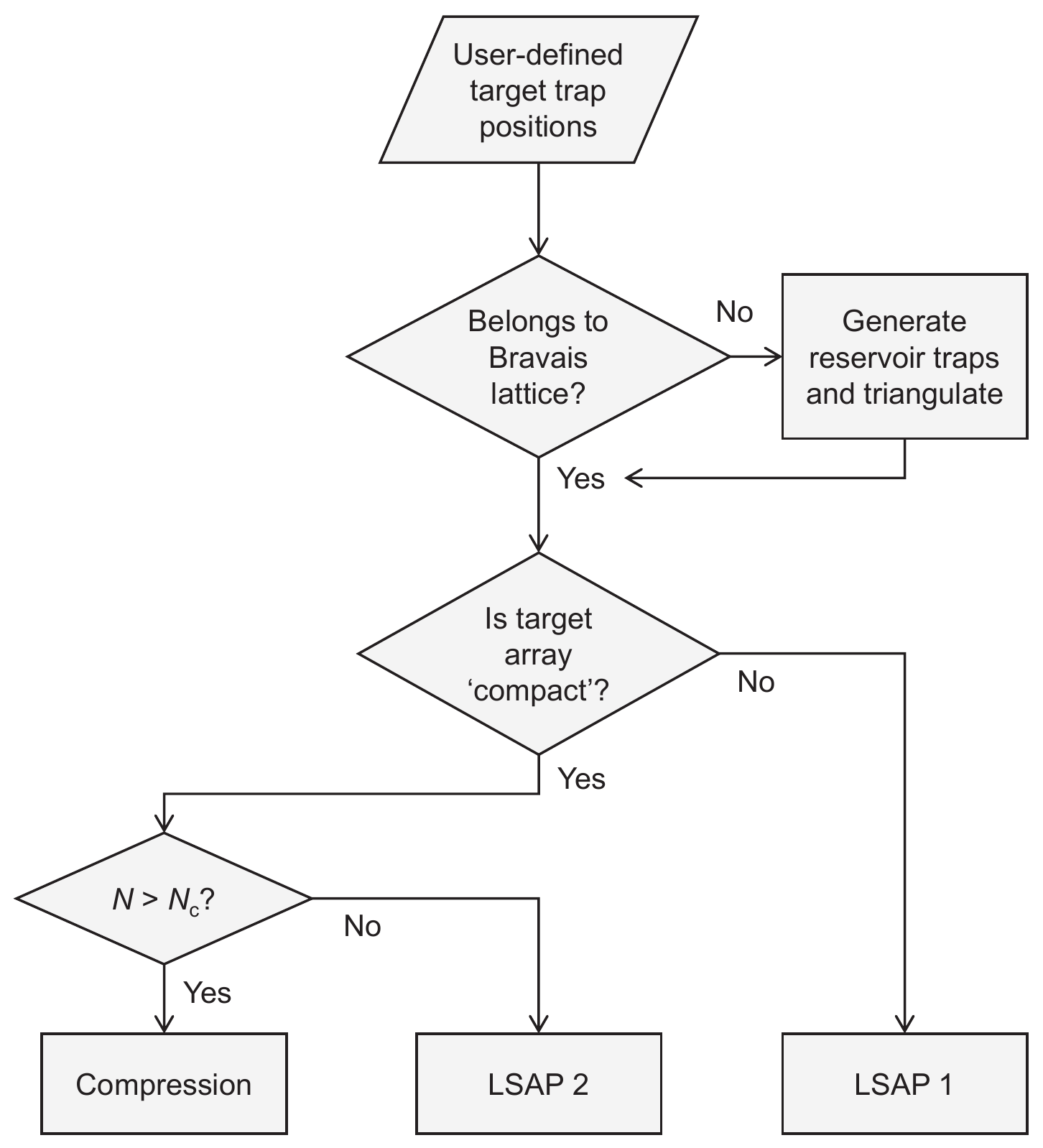}
\caption{\textbf{Algorithm choice flowchart.} The best-suited algorithm to be used depends on the characteristics of the target array.}
\label{fig:flow}
\end{figure}

The experimental setup has been described in~\cite{Barredo2016}. Using an SLM (Hammamatsu X10468-02), a fixed pattern of optical dipole traps at \SI{850}{\nano\meter} is generated in the focal plane of a high-numerical aperture ($NA = 0.5$) aspheric lens. With an available laser power of \SI{\sim 1}{\watt}, we can generate up to 200 traps with a $1/e^2$-radius of \SI{\sim 1}{\micro\meter} and a typical trap depth of \SI{\sim 1}{\milli\K}, resulting in a radial (longitudinal) trapping frequency around \SI{100}{\kilo\Hz} (\SI{20}{\kilo\Hz}). Initially, the traps are stochastically loaded with single atoms at a temperature of \SI{\sim 10}{\micro\K} from a magneto-optical trap of $^{87}$Rb atoms; the typical loading time is $\sim$\SI{150}{\milli\second}. An initial fluorescence image (\SI{20}{\milli\second}) determines the initial occupancy of the traps, which is 50 to 60~$\%$ on average.

To assemble a target array, we use a single \SI{850}{\nano\meter} dipole trap with a $1/e^2$-radius of \SI{\sim 1.3}{\micro\meter}, steered by a 2D-AOD, which can pick-up an atom from a static trap by ramping up its depth to \SI{\sim 10}{\milli\K}, subsequently moving and then releasing the atom at the position of an empty static trap. After the assembly, a fluorescence image with an exposure time of \SI{20}{\milli\second} determines the occupancy of the target array, before we perform an actual experiment, e.g. quantum simulation of a spin model, by exciting the atoms to Rydberg levels~\cite{Browaeys2020}. This technique allows us to perform experiments with a typical repetition rate of \SI{\sim 3}{\Hz}. 

Once the trap array has been generated, we equalize the trap intensities using the fluorescence signal of the loaded traps \cite{note-Equalization-Scheme}. Then, the choice of the optimal algorithm to be used for assembly, among the three ones described above, is made according to the characteristics of the target array to assemble, as described in Fig.~\ref{fig:flow}. 

\begin{figure}[t]
	\centering
	\includegraphics[width=85mm]{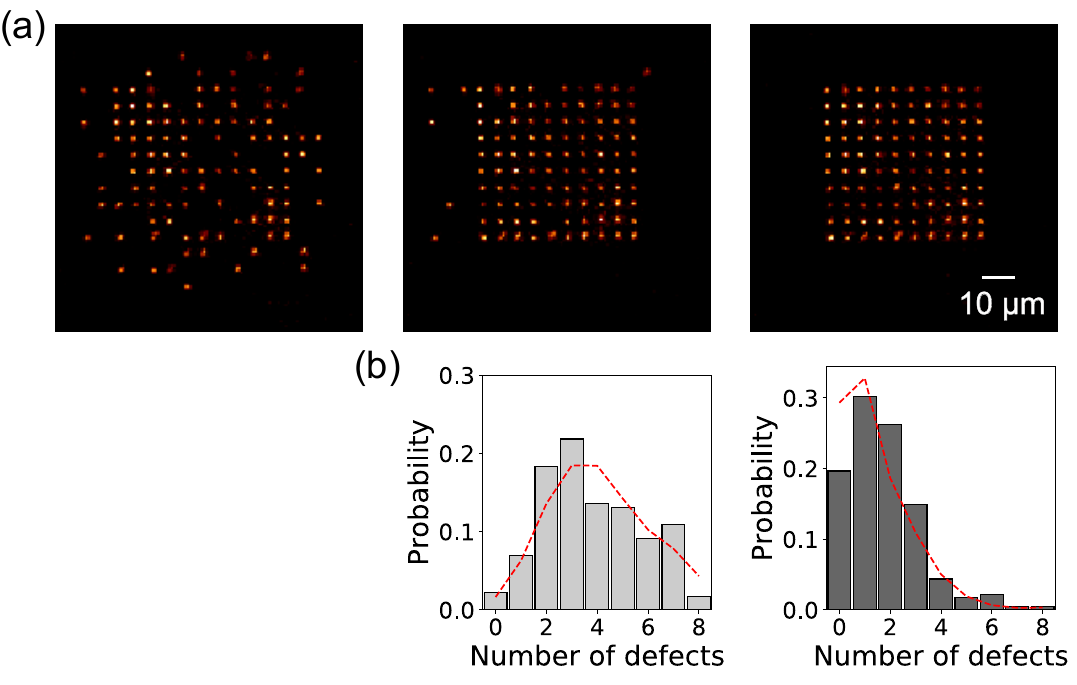}
	\caption{\textbf{Multiple rearrangement cycles.} The probability to assemble a defect-free array can be increased by starting with more than $N$ atoms and repeating the rearrangement cycle more than once. (a) on the shown $10\times10$ compact target square array, we can increase the probability to create a defect-free array by a factor 10 (from 2\% to 20\%), when starting with 225 atoms and performing a second cycle. (b) A Monte-Carlo simulation (red) of the first cycle and second cycle, including the measured efficiencies of performing the moves and vacuum lifetime, reproduces the experimental distribution of defects reasonably well. }
	\label{fig:2ndRearrangement}
\end{figure}

Finally, in order to further improve the success probability of assembling a defect-free array, we apply multiple rearrangement cycles (similar to \cite{Endres2016,deMello2019}). At the end of the first rearrangement process, we keep the excess atoms and determine the defects with a fluorescence image. We then fill these defects (Fig.~\ref{fig:2ndRearrangement}(a)). This process can be repeated until a defect-free array is obtained, and excess atoms are removed. However, since this procedure requires more than $N$ initial atoms, a high efficiency of a single rearrangement cycle is still essential as laser power is a limiting factor for scaling up the number of atoms. Figure~\ref{fig:2ndRearrangement}(b) shows the probability distribution of the number of defects (missing atoms) after a single (left) or two (right) rearrangement cycles, showing the benefit of performing several cycles. 

\begin{figure*}
	\centering
	\includegraphics[width=0.9\textwidth]{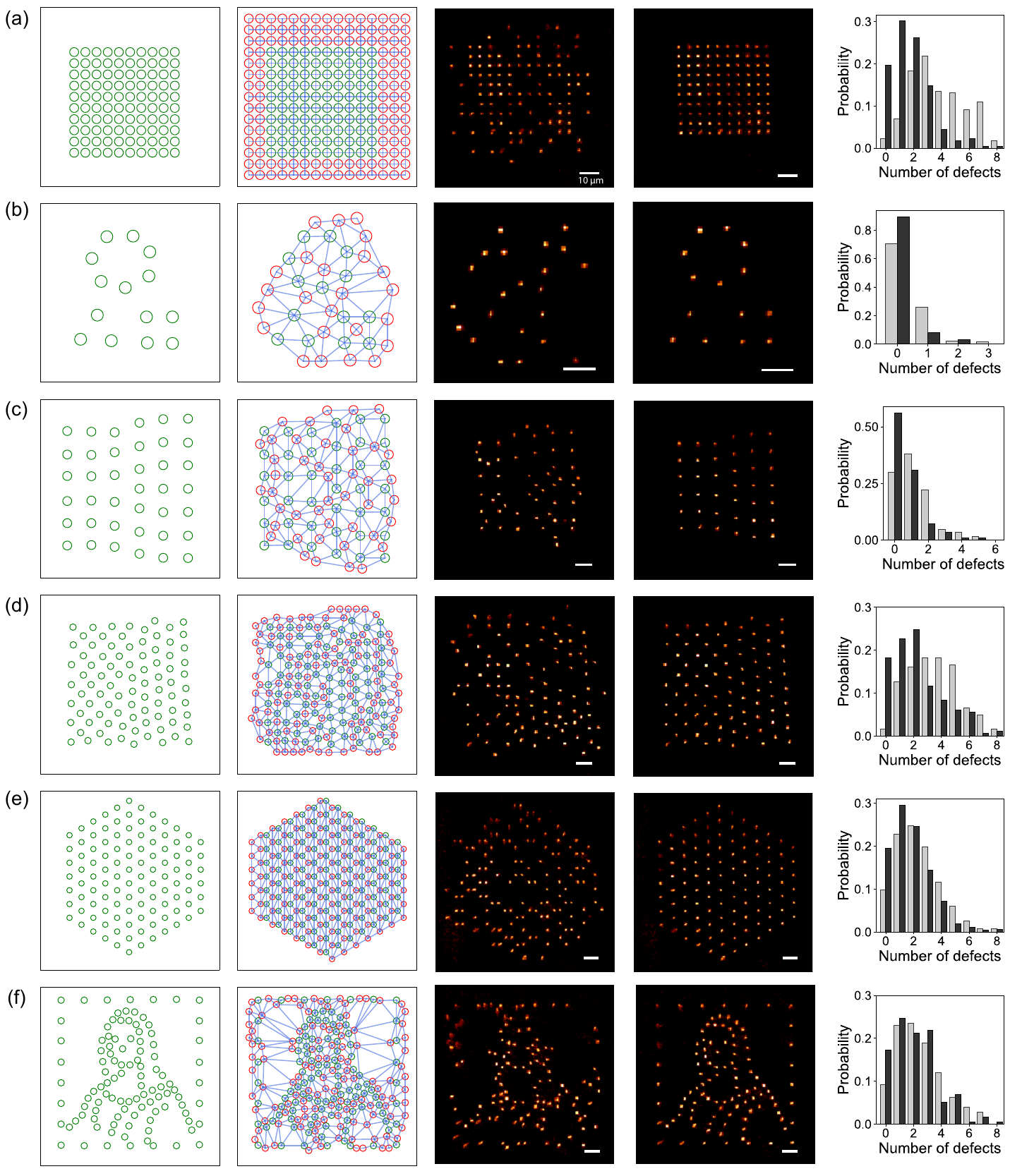}
	\caption{\textbf{Gallery of assembled arbitrary structures.} From left to right: target structure, structure with the generated reservoir traps (in red) and the allowed paths connecting traps, fluorescence image of an initial random loading, fluorescence image of the assembled structure, probability distribution of the number of defects after a rearrangement cycle (gray) and after two such cycles (dark gray). All white scale bars are \SI{10}{\micro\meter}. (a) a compact square array ($N=100$); (b) the arbitrary array used as an example in Sec.~\ref{sec:arb} ($N=14$); (c) an edge dislocation in a square lattice ($N=39$); (d) a grain boundary between a square and a triangular lattice ($N=91$); (e) a patch of a triangular lattice ($N=108$); (f) an atomic rendering of Mona Lisa ($N=106$).  }
	\label{fig:Gallery}
\end{figure*}

Examples of assembled structures of various types, with up to $N=108$ atoms, can be seen in Fig.~\ref{fig:Gallery}. The probability to have a given number of defects  in the final array is shown in the histograms on the right, for a single rearrangement (gray), and for two cycles (dark gray). In the latter case, even for $N>100$, defect-free arrays are obtained in about 20\% of the shots. Using a trapping wavelength closer to resonance (820~nm) in order to generate more traps for a given laser power, we have been able to assemble arrays of up to 209 atoms without any given defects.   

\section{Conclusion}

In this paper, we have shown how, without any change in the hardware used in \cite{Barredo2016}, improved algorithms can significantly improve the capabilities of a moving-tweezers atom-by-atom assembler, both in terms of possible array geometries, and in terms of achievable atom numbers thanks to the fact that fewer moves are required.  

A natural extension of this study, that we leave for future work, is to use multiple tweezers working in parallel, in the spirit of \cite{Endres2016}. This approach should be particularly easy to adapt to the compression algorithm for assembling compact, regular structures; then, assuming that the laser power for generating the multiple tweezers is not a limit, the assembly time could scale as $\sqrt{N}$, making it possible to assemble structures with several hundreds of atoms. Combined with other technical improvements, using e.g. cryogenic environments to drastically extend  the vacuum-limited lifetime, reaching a scale of a thousand atoms or more thus seems realistic in a relatively near future, which would open up a variety of exciting applications in quantum science and technology. 

\begin{acknowledgments}
We thank Lo\"ic Henriet and Henrique Silv\'erio for useful discussions, and Gilles Kuhorn for contributions to testing some of the algorithms.  This project has received funding from the European Union's Horizon 2020 research and innovation program under grant agreement no. 817482 (PASQuanS), and from the R{\'e}gion {\^I}le-de-France in the framework of DIM SIRTEQ (project CARAQUES). \end{acknowledgments}


\begin{thebibliography}{30}
\bibitem [{not({\natexlab{a}})}]{note-LSAP-Scaling}%
  \BibitemOpen 
  \bibinfo {note} {{In the worst-case, the
  Hungarian matching algorithm is known to scale as $N^3$, however we observe
  empirically that for the current problem, and for the values of $N$ up to a
  few hundreds considered here, the average runtime of our LSAP and reordering
  algorithm scales roughly  as $N^2$.}}\BibitemShut {Stop}%
\bibitem [{not({\natexlab{a}})}]{note-Lookup-table}%
\BibitemOpen 
\bibinfo {note} {{To reduce the computation time during the experiment, we precalculate a look-up table with the shortest paths and path lengths between all trap pairs. During each assembly cycle, the cost matrix for the LSAP algorithm is found as a submatrix of the look-up table.}}\BibitemShut {Stop}%
\bibitem [{not({\natexlab{b}})}]{note-Equalization-Scheme}%
  \BibitemOpen
  \bibinfo {note} {{It is of importance
  that all micro-traps have a good optical quality, and in particular the same
  depth such that (i) single-atom loading does indeed occur with a probability
  of $\sim 1/2$; (ii) the fluorescence signal from any given trap allows for
  efficient identification of the presence of a single atom. We now equalize
  the trap depths by a direct optimization of the fluorescence time trace of
  each single trap, altering the trap intensity until we fulfill criteria (i)
  and (ii). This procedure will be described in detail in future
  work.}}\BibitemShut {Stop}%
\end{thebibliography}
\end{document}